\documentclass{desyproc}

\usepackage{lineno}

\begin{document}
%------------------------------------
\title{ATLAS results on soft diffraction}
%for single authors the superscripts are optional
\author{{\slshape Simone Monzani$^1$}\\[1ex]
$^1$Universit\'a di Bologna, Via Irnerio 46, I-40127 Bologna, Italy\\
On behalf of the ATLAS Collaboration}

% if the proceedings are available online (e.g. at Indico)
% please enter the contribution ID or file_name below for the DOI
%\contribID{32}
\contribID{smith\_joe}

% TO THE CONFERENCE EDITORS:
% please update the following information
% before sending the template to the authors
% \confID{800}  % if the conference is on Indico uncomment this line

\acronym{EDS'13} % if you want the Acronym in the page footer uncomment this line

\maketitle

%\linenumbers

\begin{abstract}
%One of the main purposes of the studies on diffraction physics is the determination of the total inelastic cross section in pp collisions.
The measurements of the total inelastic cross section and the differential inelastic cross section as a function of rapidity gap are presented. The data used for these studies were collected in $pp$ collisions at a center-of-mass energy of 7 TeV with the ATLAS detector at the LHC in 2010.
%Two measurements of the inelastic cross section with a center of mass energy of 7 TeV with the ATLAS detector at LHC, are here presented.\\
%The first is a total cross section, the second a differential one as function of rapidity gap.
\end{abstract}

%\section{The ATLAS detector}

%A detailed description of the ATLAS detector can be find elsewhere \cite{ATLAS}. The main components used in the studies presented in this report are summarized here. The beam-line is surrounded by the 'inner detector' tracking system, which covers the pseudorapidity
%range $|\eta|<2.5$.\footnote{ The pseudorapidity is defined in terms of the polar angle $\theta$ as $\eta = - ln$ $tan$$(\theta/2)$}
%This detector consists of silicon pixels, silicon strips and straw tube detectors and is enclosed within a uniform 2T solenoidal magnetic field.
%The calorimeters lie outside the tracking system. The
%highly segmented electromagnetic, the hadronic liquid argon
%sampling calorimeters and the hadronic tile calorimeter cover the range $|\eta|<4.9$.\\
%Minimum bias trigger scintillator (MBTS) detectors
%are mounted in front of the end-cap calorimeters on
%both sides of the interaction point and cover the pseudorapidity range $2.1<|\eta|<3.8$.

%\begin{wraptable}{r}{0.45\textwidth}
%\centerline{\begin{tabular}{|l|r|}
%\hline
%Column1            & Column2 \\
%\hline
%\end{tabular}}
%\caption{Example table}
%\label{tab:limits}
%\end{wraptable}

\section{Introduction}

This report presents the measurements of the total and differential inelastic cross section in $pp$ collisions at a center-of-mass energy ($\mathrm{\sqrt{s}}$) of 7 TeV with the ATLAS detector \cite{ATLAS} at the LHC. The total $pp$ cross section has two components: the elastic and the inelastic one. This latter, in turn, is composed of 4 terms: non-diffractive (ND), single diffractive (SD), double diffractive (DD) and central diffractive (CD).
Single diffraction is a process in which one of the two protons dissociates in a final state X and the other stays intact; in double diffraction processes both protons dissociate into two final states X and Y. In single and double diffractive inelastic scattering a colourless Pomeron exchange leads to a rapidity gap, a region in rapidity devoid of particles. Finally in central diffraction processes none of the protons dissociate but a final state centrally produced arises.\\
Pythia6 \cite{Pythia6}, Pythia8 \cite{Pythia8} and PHOJET \cite{PHOJET} generators are used to predict the properties of the inelastic collisions.\\
Experimentally the diffractive processes are described using the kinematic variable $\mathrm{\xi_{X,Y}=\frac{{M_{X,Y}}^{2}}{s}}$, where $\mathrm{M_{X,Y}}$ is the dissociative mass of the final state X or Y and s already introduced.
%\begin{figure}[hb]
%\centerline{\includegraphics[width=0.45\textwidth]{SD}}
%\centerline{\includegraphics[width=0.45\textwidth]{DD}}
%\centerline{\includegraphics[width=0.45\textwidth]{CD}}
%\caption{Figure title.}\label{Diffraction}
%\end{figure}

\section{Total inelastic cross section}

\begin{figure}[hb]
\centerline{\includegraphics[width=0.45\textwidth]{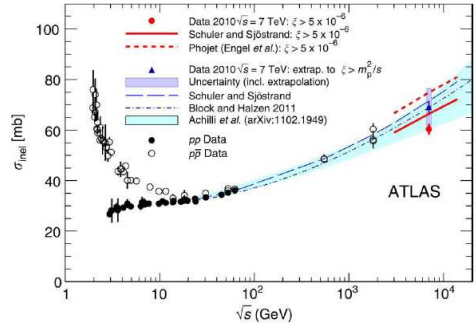}}
\caption{Total inelastic cross section as function of $\sqrt{s}$ \cite{TCS}.}\label{TI}
\end{figure}

The measurement of the total $pp$ inelastic cross section was performed using data collected in one run in March 2010 at $\mathrm{\sqrt{s}=7}$ TeV corresponding to an integrated luminosity of 20 $\mu$b$^{-1}$, with a mean number of collisions per bunch crossing of about 0.01.\\
For triggering the Minimum Bias Trigger Scintilator (MBTS) was used. The MBTS system is composed by a set of scintillators located in front of the end-cap calorimeters of the ATLAS detector on both sides of the interaction point and covers the pseudorapidity range $\mathrm{2.1<|\eta|<3.8}$. The MBTS trigger is used to define two samples: single-sided events, with signal present exclusively in one side, and inclusive events, with signal in one or both sides of the detector.\\
The total inelastic cross section was obtained using the formula %$\sigma_{inel}(\xi>10^{-6})=\frac{(N-N_{BG})}{\varepsilon_{t}\cdot L_{Int}}\cdot\frac{1-f_{\xi<10^{-6}}}{\varepsilon_{Sel}}$
\begin{equation}
\label{TICS}
\mathrm{\sigma_{inel}(\xi>5\cdot10^{-6})}=\frac{(\mathrm{N}-\mathrm{N_{BG}})}{\mathrm{\varepsilon_{t}}\cdot \mathrm{L_{Int}}}\cdot\frac{1-\mathrm{f_{\xi<5\cdot10^{-6}}}}{\mathrm{\varepsilon_{Sel}}}
\end{equation}
where N and N$_{BG}$ are the number of selected and background events. The background is mainly due to beam-related interactions and was determined using the number of
events selected in correspondence with non-colliding bunches. $\mathrm{L_{Int}}$ is the integrated luminosity, $\mathrm{\varepsilon_{t}}$ the trigger efficiency (99.98 $\%$), $\mathrm{\varepsilon_{Sel}}$ the selection efficiency (98.77 $\%$) and $\mathrm{f_{\xi<5\cdot10^{-6}}}$ (0.96 $\%$) the fraction of events with $\mathrm{\xi<5\cdot10^{-6}}$, evaluated using the default Donnachie and Landshoff (DL) model \cite{DL} as implemented in Pythia 8. The MBTS acceptance was found using MC and it is used to define the fiducial region of the measurement, corresponding approximately to $\mathrm{\xi>5\cdot10^{-6}}$, that is to M$_{X,Y}$ greater than 15.7 GeV.\\
The fractional contribution of diffractive events, $\mathrm{f_{D}=\frac{\sigma_{SD}+\sigma_{DD}+\sigma_{CD}}{\sigma_{Inel}}}$, is constrained by
R$_{SS}$%=\frac{N_{SS}}{N_{Inc}}$
, the ratio of single-sided to inclusive events.
%\begin{equation}
%
%\label{f}
%f_{D}=\frac{\sigma_{SD}+\sigma_{DD}+\sigma_{CD}}{\sigma_{Inel}}
%\end{equation}
%\begin{equation}
%
%\label{R}
%R_{SS}=\frac{N_{SS}}{N_{Inc}}
%\end{equation}
%where $N_{SS}$ and $N_{Inc}$ are respectively the single sided and inclusive events.
%The $\varepsilon_{Sel}$ and $f_{\xi<5\cdot10^{-6}}$ factors are taken from the tuned MC simulation. In order to reduce the uncertainties in the factors taken from the MC simulation the $f_{D}$ parameter for each used generator is constrained.
R$_{SS}$ was measured to be $\mathrm{[10.02\pm0.03 (stat.)^{+0.1}_{-0.4}}$ $\mathrm{(syst.)}]\%$ which, using the default
DL model, corresponds to $\mathrm{f_{D}=26.9^{+2.5}_{-1.0}\%}$.
%$R_{SS}$ was obtained by ATLAS, and its measured value is around 10 $\%$  OK
%corresponding to $f_{D}\sim 23 \%$.    OK
In Fig. \ref{TI} the total inelastic cross section is presented as a function of $\sqrt{s}$. The ATLAS measurement obtained through the equation above is represented by the red circle and is equal to $\mathrm{60.3\pm0.5(sys.)\pm2.1(lumi)}$ mb. The phenomenological models underestimate the observed result. The extrapolated cross section (represented by the blue triangle), obtained extending the M$_{X,Y}$ range to the proton mass, was evaluated to be $\mathrm{69.4\pm2.4(exp.)\pm6.9(extr.)}$ mb, consistent with the phenomenological predictions. Black and white dots refers to previous $pp$ and $p\overline{p}$ data respectively \cite{TCS}.

\section{Inelastic cross section as a function of rapidity gap}

\begin{figure}[hb]
%\subfloat[\label{TICS}]{
%\includegraphics[width=0.45\textwidth]{TICSP}}
\includegraphics[width=0.45\textwidth]{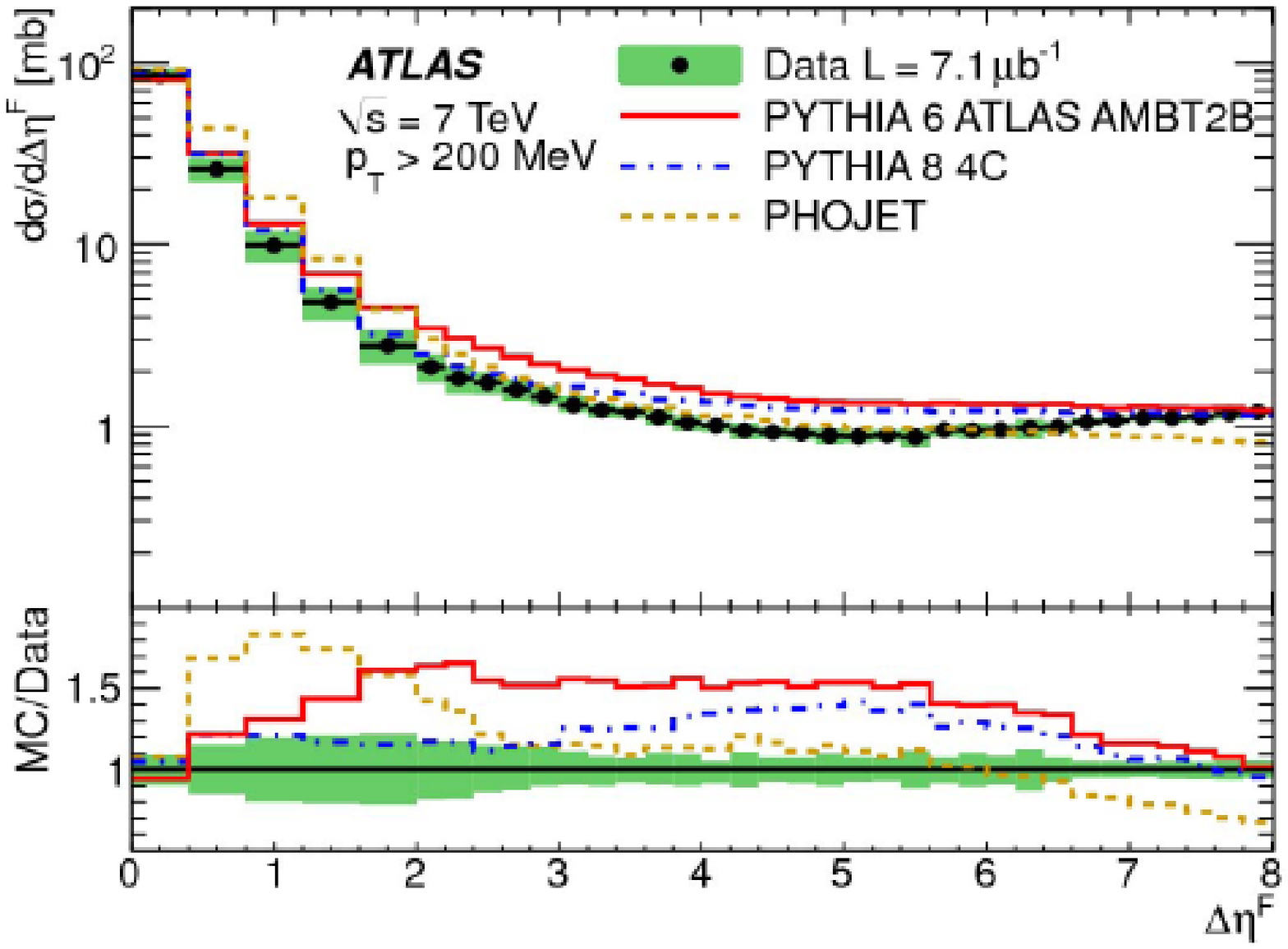}
\includegraphics[width=0.45\textwidth]{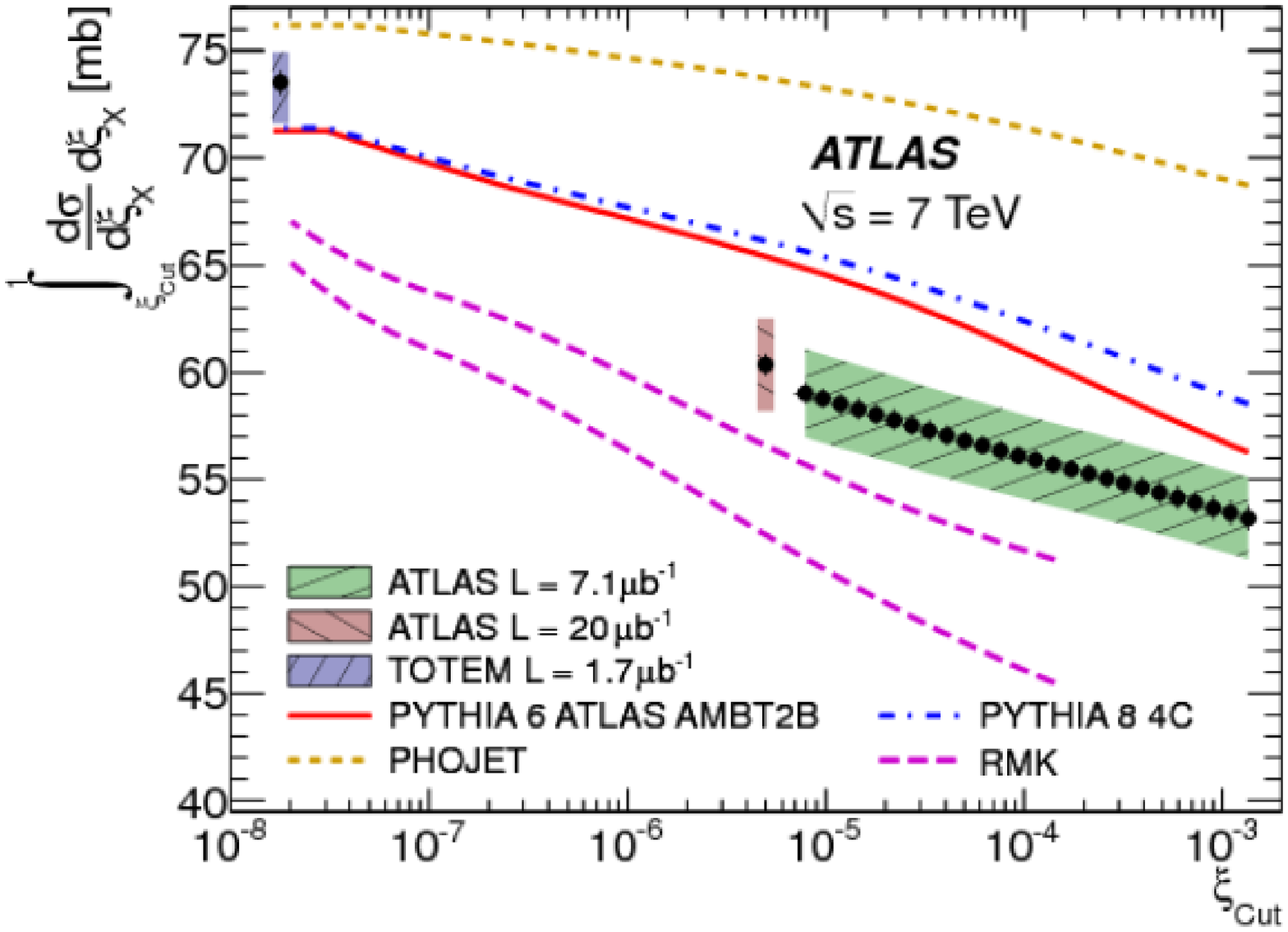}
\caption{Differential inelastic cross section as a function of forward rapidity gap (a) and $\mathrm{\xi_{CUT}}$ (b) \cite{RG}.}\label{DICSTP}
\end{figure}

The differential inelastic cross section measurement was performed using data corresponding to an integrated luminosity of 7 $\mu$b$^{-1}$ collected with a negligible rate of multiple proton-proton interactions. The MBTS trigger was used for the online event selection.\\
The forward rapidity gap ($\mathrm{\Delta \eta_{F}}$) is defined as the larger of the two empty pseudorapidity regions
extending between the edges of the detector acceptance
%at $\eta$= 4.9 or $\eta$=-4.9
and the closest track or calorimeter cluster. On the particle level, it is defined without stable charged and neutral particles with $\mathrm{p_{T}>200}$ MeV.
A the differential inelastic cross section as a function of rapidity gap is shown in Fig \ref{DICSTP}.
The plot shows an exponential decrease at low $\mathrm{\Delta \eta_{F}}$ due to the non-diffractive events
%\begin{figure}[hb]
%\centerline{\includegraphics[width=0.45\textwidth]{DICSP}}
%\caption{Differential inelastic cross section as function of rapidity gap}\label{DI}
%\end{figure}
while, for $\mathrm{\Delta \eta_{F}>3}$, a flat behavior is observed as predicted by the theory ($\Delta\eta$ proportional to -log $\xi$).\\
All the MC models considered for the comparison lie above data, Pythia8 is closer than other models to the data at low $\mathrm{\Delta \eta_{F}}$,
while PHOJET is better at intermediate values of $\mathrm{\Delta \eta_{F}}$.\\
%Assuming the triple Pomeron phenomenology, the differential inelastic cross section is exponentially proportional to an $\alpha(t)$
%term known as Pomeron trajectory that is parametrised as $\alpha (0)+\alpha't$. Pythia8 was run with different values of the Pomeron intercept $\alpha(0)$
%obtaining different distributions for the differential inelastic cross section as function of $\Delta \eta_{F}$. The $\alpha(0)$ which
%best describes the data (the one with the best $\chi^{2}$ from a fit to the data for $\Delta \eta_{F}>6$)
%was extracted giving a value of $1.058\pm0.003(stat.)^{0.034}_{-0.039}(sys.)$ for the Pomeron intercept.\\
The differential inelastic cross section was also measured as a function of $\mathrm{\xi_{CUT}}$ (requiring $\mathrm{\xi_{X,Y}>\xi_{CUT}}$). %that is equivalent to measure it as a function of $\Delta \eta_{F,CUT}$ ($\Delta \eta_{F,CUT}$ is the maximum value of $\Delta \eta_{F}$), since the two values are correlated.
In Fig \ref{DICSTP}b, different measurements of the integrated cross section as a function of $\mathrm{\xi_{CUT}}$ from TOTEM and ATLAS are shown together with MC predictions. The most recent results from ATLAS was obtained for different values of integration limit and are in agreement with the previous ATLAS result \cite{RG}.

%\begin{figure}[hb]
%\centerline{\includegraphics[width=0.45\textwidth]{BackgroundComparisons2.eps}}
%\caption{Figure title.}\label{FigureLabel}
%\end{figure}

\section{Conclusions}

ATLAS measurements of the total and differential $pp$ inelastic cross section at $\sqrt{s}=7$ TeV have been presented.
The total inelastic cross section for $\mathrm{\xi>5\cdot10^{-6}}$, within the detector acceptance, was measured to be
$\mathrm{60.3\pm0.5(sys.)\pm2.1(lumi)}$ mb while the extrapolation to the full phase space gives a value of $\mathrm{69.4\pm2.4(exp.)\pm6.9(extr.)}$ mb.
The differential inelastic cross section as a function of the rapidity gap was also measured. All the considered models overestimate the data and none of them gives
a detailed description of the shape of the distribution.

%Figure \ref{FigureLabel} shows an example of a figure and related
%caption. This is how you reference an article~\cite{H1}.

%\section{Acknowledgments}

%To acknowledge funding bodies etc., a special section may be placed
%before the bibliography.

% ****************************************************************************
% BIBLIOGRAPHY AREA
% ****************************************************************************

\begin{footnotesize}
% IF YOU DO NOT USE BIBTEX, USE THE FOLLOWING SAMPLE SCHEME FOR THE REFERENCES
% ----------------------------------------------------------------------------

\end{footnotesize}
\end{document}